\documentclass[onecolumn,tightenlines,showpacs,nofootinbib,preprintnumbers,
amsmath,amssymb]{revtex4}

\usepackage[usenames,dvipsnames]{color}

\usepackage{graphicx}
\usepackage{bm}
\usepackage{lscape}

\usepackage{setspace}
\begin{document}
\begin{spacing}{1.5}


\title{Effects of Shock Waves on Neutrino Oscillations in Three Supernova Models}

\author{Jing Xu$^{1}$\footnote{Email: xj2012@mail.bnu.edu.cn}, Li-Jun Hu $^{1}$\footnote{Email: hulj@mail.bnu.edu.cn}, Rui-Cheng Li$^{1}$\footnote{Email: Rui-cheng Li@163.com}, Xin-Heng Guo$^{1}$\footnote{Corresponding author, Email: xhguo@bnu.edu.cn} and Bing-Lin Young$^{2,3}$\footnote{Email: young@iastate.edu}}
\affiliation{\small $^{1}$College of Nuclear Science and
Technology, Beijing Normal University, Beijing 100875, China\\
\small $^{2}$Department of Physics and Astronomy, Iowa State
University, Ames, Iowa 5001, USA\\
$^{3}$Institute of Theoretical Physics, Chinese Academy of Sciences, Beijing 100190, China }

\begin{abstract}

It has been realized that the shock wave effects play an important role in neutrino oscillations during the supernova explosion. In recent years, with the development of simulations about supernova explosion, we have a better understanding about the density profiles and the shock waves in supernovae than before. It has been shown that the appearance of shock waves not only varies with time, but is also affected by the mass of the supernova. When the mass of the supernova happens to be in a certain range (e.g. it equals 10.8 times the mass of the sun), there might be a reverse shock wave, another sudden change of density except the forward shock wave, emerging in the supernova. In addition, there are some other time-dependent changes of density profiles in different supernova models. Because of these complex density profiles, the expression of the crossing probability at the high resonance, $P_H$, which we used previously would be no longer applicable. In order to get more accurate and reasonable results, we use the data of density profiles in three different supernova models obtained from simulations to study the variations of $P_s$ (the survival probability of $\nu_e\rightarrow\nu_e$), as well as $P_c$ (the conversion probability of $\nu_x\rightarrow\nu_e$). It is found that the mass of the supernova does make a difference on the behavior of $P_s$, and affects $P_c$ at the same time. With the results of $P_s$ and $P_c$, we can estimate the number of $\nu_e$ remained after they go through the matter in the supernova.

\end{abstract}


\maketitle

\section{Introduction}

Since 1980's, supernova neutrinos have been a focus of our attention for a long time \cite{Arnett1}-\cite{Bethe1}. In recent years, the issues about neutrino detection experiments and simulations of supernova explosion have always been hotspots in scientific research fields \cite{Ardellier}-\cite{Janka0}. Thanks to the results of simulations and development of theories on supernovae, the whole process of supernova explosion has been understood much better now \cite{Bruenn}-\cite{Kotake}. The post bounce period, one stage in the process of supernova explosion, can be divided generally into the accretion phase and the cooling phase. During the supernova explosion, both the collective effects and the shock wave effects play important roles in neutrino oscillations \cite{Hudepohl}\cite{Chakraborty0}. As we know, in the post bounce period, a large number of supernova neutrinos go through the supernova matter, carrying an enormous amount of energy, and emit from the supernova \cite{Spergel}\cite{Giunti}. Meanwhile, the spreading outward energy makes significant changes to the supernova matter and arouses the shock wave, which has an impact on neutrino oscillations in the supernova \cite{Fogli1}-\cite{Schirato}.

The essence of the shock wave effect is the Mikheyev-Smirnov-Wolfenstein (MSW) effect, caused by the interactions of neutrinos with matter and dominated by the density profile of the supernova as well as neutrino mixing angles \cite{Takahashi1}-\cite{Kuo}. The neutrino crossing probability in the high resonance, $P_H$, will experience complex changes resulting from the high-density and unstable matter in the supernova \cite{Kuo}-\cite{Kachelriess}. In the past, the simplified model of supernova density profile, which only contains the forward shock wave, was used to study the behavior of $P_H$ \cite{Fogli2}. With this method, we could roughly figure out the impact of the shock wave effects on neutrino oscillations in the supernova \cite{Guo}-\cite{Xu}. Afterwards, it was found that besides the forward shock wave there would be a reverse shock wave and some turbulence existing in the supernova during the process of explosion, all of which have influences on neutrino oscillations \cite{Kneller}-\cite{Couch}. Presently, from the results of simulations aiming at different supernova models, it is known that the actual density profiles in supernova are distinctly different from one to another due to the difference of their masses \cite{Lund}\cite{Wongwathanarat}. For this reason, the original simplified model of the density profile cannot provide the images of time-varying densities in different supernova models comprehensively. Therefore, one has to find out the images in other ways. One possible way is to get new simplified density profiles according to the simulation results of different supernova models, respectively. In this way, it is necessary to do data analysis and fitting for densities, which is beyond the scope of the present work. Therefore, we decide to use the simulation results of supernova density profiles to discuss the variations of $P_s$ (the survival probability of $\nu_e\rightarrow\nu_e$) and $P_c$ (the conversion probability of $\nu_x\rightarrow\nu_e$, $\nu_x$ represents $\nu_{\mu}$ and $\nu_{\tau}$) instead of applying the simplified model of density profiles to study the behavior of $P_H$.

This paper is divided into five parts. In Section 2, we give a brief review about neutrino oscillations in matter. In Section 3, we use the simulation results of three supernova models to give the images of time-varying density profiles in order to find the similarities and differences among them. In Section 4, based on the theory in Section 2 and using the density data in Section 3, we calculate $P_s$, $P_c$ and the ratio of $\nu_e$ remained after neutrinos go through the matter in three supernova models. In addition, we compare the diverse appearances of shock waves among these supernova models and extract how the time-dependent matter affects neutrino oscillations in supernovae. The last section consists of a summery and some discussions of this paper.

\section{Neutrino oscillations in matter}

Neutrinos might be regarded as either a Dirac or a Majorana particle. No matter which kind of particle it is, eliminating the spin structure from its propagation equation yields the Klein-Gordon equation \cite{Kuo}. In the flavor basis, different from the case in vacuum, the propagation equation of neutrino in matter has a potential energy item, which stems from the interactions between neutrinos and matter,
\begin{eqnarray}
-i\frac{d}{dt}\left|{{\nu}_{\alpha}}\right\rangle&=&\frac{1}{2p}M^{2}\left|{{\nu}_{\alpha}}\right\rangle,
\label{n}\\
M^{2}&=&\left[U\left[\begin{array}{cc}m_{1}^{2}&0\\0&m_{2}^{2}\\\end{array}\right]U^{\dagger}+\left[\begin{array}{cc}A&0\\0&0\\\end{array}\right]\right]\nonumber.
\end{eqnarray}

In Eq. (\ref{n}), ${\nu}_{\alpha}$ stands for neutrinos in flavor eigenstate, $U$ the unitary transformation between the flavor and mass bases, $m_{1}^{2}$ and $m_{2}^{2}$ the square of mass eigenvalues, and $A$ is the parameter containing the effect of matter on neutrino propagation, which is give by
\begin{eqnarray}
A=2Vp=2\sqrt{2}G_{F}N_{e}p, \label{A1}
\end{eqnarray}
where $G_{F}$ is the Fermi constant, $N_{e}$ the number of electrons, and $p$ the momentum of neutrino. For antineutrinos, $A{\rightarrow}-A$ and $U{\rightarrow}U^{*}$.

For neutrinos, $p\sim E$, so the expression of $A$ can be written as
\begin{eqnarray}
A=2\sqrt{2}G_{F}N_{e}E=2\sqrt{2}G_{F}(Y_{e}/m_n)\rho E, \label{A2}
\end{eqnarray}
where $\rho$ is the matter density in the supernova, $Y_e$ the number of electron per nucleon, and $m_n$ the nucleon mass.

In the study of supernova neutrinos, we work in the approximation that there are only two neutrino flavors, $\nu_{e}$ and $\nu_{x}$ ($\nu_{x}$ represents $\nu_{\mu}$ and $\nu_{\tau}$) \cite{Kuo}. Acting like an induced mass squared of the electron neutrino, $A$ arises from the propagation of the electron neutrino through a background of electrons. With $c=1$ and $t\rightarrow x$, the equation of two flavors in matter is written as
\begin{eqnarray}
i\frac{d}{dx}\left[\begin{array}{c}{\nu}_{e}\\{\nu}_{x}\\\end{array}\right]&
=&\frac{1}{2E}M^{2}\left[\begin{array}{c}{\nu}_{e}\\{\nu}_{x}\\\end{array}\right]\nonumber\\&
=&\frac{1}{2E}\left[U\left[\begin{array}{cc}m_{1}^{2}&0\\0&m_{2}^{2}\\\end{array}\right]U^{\dagger}
+\left[\begin{array}{cc}A&0\\0&0\\\end{array}\right]\right]
\left[\begin{array}{c}{\nu}_{e}\\{\nu}_{x}\\\end{array}\right]\label{n2}\\
&=&\frac{1}{4E}\left[(\Sigma+A)+\left[\begin{array}{cc}{A-{\Delta}C_{2\theta}}&{\Delta}S_{2\theta}\\{\Delta}S_{2\theta}&{-A+{\Delta}C_{2\theta}}\\\end{array}\right]\right]
\left[\begin{array}{c}{\nu}_{e}\\{\nu}_{x}\\\end{array}\right],\nonumber
\end{eqnarray}
where $\Sigma=m_{2}^{2}+m_{1}^{2}$, $\Delta=m_{2}^{2}-m_{1}^{2}$, the mixing angle $\theta$ refers to $\theta_{13}$ and $C_{2\theta}=\cos{2\theta}$, $S_{2\theta}=\sin{2\theta}$. In recent years, the new data about $\theta_{13}$ have been provided by some neutrino experiments \cite{Ardellier}\cite{Abe1}-\cite{An1}\cite{Abe2}\cite{Ahn}. In the following, we will use $\theta_{13}\simeq8.8^{\circ}$ provided by the Daya Bay Collaboration \cite{An1}.

By solving Eq. (4), we can get how $\nu_e$ and $\nu_x$ evolute with distance, then the survival probability and conversion probability are given as follows:
\begin{eqnarray}
P_s=P({\nu_e} \rightarrow {\nu_e})={\left|\left\langle{{\nu}_{e}}|{{\nu}_{e}}\right\rangle\right|}^2, \nonumber\\
P_c=P({\nu_x} \rightarrow {\nu_e})={\left|\left\langle{{\nu}_{e}}|{{\nu}_{x}}\right\rangle\right|}^2.
\end{eqnarray}

\section{The density profiles of different SN models}

In the previous study, we used the simplified density profile to represent the real state of density without considering the mass of the supernova \cite{Guo}-\cite{Xu}. In this section, based on simulation data we can provide the density profiles at different times in three supernova models. These three supernova models are 8.8, 10.8 and 18 times the mass of the sun, respectively, so they will be called as the 8.8$M_{\odot}$ model, the 10.8$M_{\odot}$ model and the 18$M_{\odot}$ model respectively for simplicity in the following.

\subsection{The 8.8$M_{\odot}$ model}

In Fig. 1 it is shown how the density changes with the radius within 6 seconds after the supernova explosion in the 8.8$M_{\odot}$ model.

It can be seen that on each density curve there is an obvious decrease at about $10^{1.5}$ km of radius, and this decline gets more and more notable with the increase of time. For example, at the time of 0.2 s, the density declines about $10^{5}$ g/cm$^{3}$ from 10 km to $10^{2}$ km, while when $t$ = 6 s, the density declines about $10^{7}$ g/cm$^{3}$ from 10 km to $10^{1.5}$ km. After the sharp decrease, the density continues to decline, but relatively slowly. In this slow-decline stage, from $10^{2}$ km - $10^{4}$ km, the density goes down about $10^7$ g/cm$^{3}$. Except for the curve of 1 s, at the end of other curves, there is an abnormal increase of density. Compared with the rise with some twists and turns on the curves of 2 s $\sim$ 6 s, at the end of 0.2 s and 0.5 s curves, the densities ascend slight increases. Yet on the curve of 1 s, there is no obvious increase at all.

Overall, in the 8.8$M_{\odot}$ model, no matter what time it is, the density decreases with the increase of radius if we do not take the ends of the curves into account. It is worth noting that the so-called shock wave caused by the changing density does not appear in the 8.8$M_{\odot}$ model at all. This is remarkably different from the simplified density profile we used previously \cite{Fogli2}.

\subsection{The 10.8$M_{\odot}$ model}

In Fig. 2, there are 11 curves to demonstrate how the density varies with the radius at different times in the 10.8$M_{\odot}$ model.

Similar to Fig. 1, there is an abrupt density decrease at about $10^{1.5}$ km of radius on each curve in Fig. 2. Then the density continues to decline slowly in comparison with the previous sharp decrease. During this phase, it can be observed that there is a density "bulge" on each curve, which is the forward shock wave we studied before. At 0.5 s, it seems that this shock wave is a bit fuzzy, but after 1 s it becomes more and more evident and moves forward with time. On the curves of $t\geqslant$ 3 s, it can be seen that there is a density "concave" located behind the forward shock wave. That is the reverse shock wave, resulting from the supernova matter squeezing with the end of the shock wave. Just like the forward shock wave, the reverse shock wave also moves forward with time. On the curves of $t\geqslant$ 6 s, a small density "bump" appears on the forward shock wave, indicating that the changes of matter within the forward shock wave are very complicated. This phenomenon has never be mentioned before \cite{Fogli2}\cite{Schirato}\cite{Huang}.

Furthermore, it is found that all the curves in Fig. 2 converge to a line at the end of themselves. Through the curve fitting method, it is found that this converging line can be represented by $\rho=r^{-2.7}$ approximately, in close proximity to $\rho=r^{-3}$, which is known as the expression of density distribution in a stable supernova as mentioned in Refs. \cite{Schirato}\cite{Kuo}.

\subsection{The 18$M_{\odot}$ model}

Same as Fig. 2, Fig. 3 shows the density profiles at 11 different points of time in the 18$M_{\odot}$ model. Just like the two former models, the density in the 18$M_{\odot}$ model also decreases rapidly at about $10^{1.5}$ km judged from the appearance of each curve and then the density declines gradually. On the curve of $t$ = 0.5 s, it seems that the forward shock wave is forming but not complete yet. While on other curves, it can be seen clearly that there is a forward shock wave moving with time. Also at the end of all curves, the density profiles of different times converge into a line. Compared with Fig. 2, in which there is a density "concave" on the curves of $t\geqslant$ 3 s, it is found that the reverse shock wave is not formed throughout in the 18$M_{\odot}$ model.

The above three figures of density profiles have something in common. That is, at about $10^{1.5}$ km of radius, there is an obvious density decline which gets more and more abrupt with time. It might be assumed that the core of the supernova, which is incredibly dense, is collapsing more and more quickly with time.

What is more, these three figures indicate that the mass of the supernova is very closely related to the shape of shock wave. If the mass is not big enough, taking the 8.8$M_{\odot}$ model as an example, there would be no shock wave (neither the forward shock wave nor the reverse shock wave) appearing in the supernova. However, if the supernova mass is too massive, for instance, the 18$M_{\odot}$ model, the reverse shock wave would not arise in it. In other words, only when the mass of the supernova is neither too small nor too big, and within a certain range (from the three models, this range should be $8.8M_{\odot}<$ the mass of SN $<18M_{\odot}$), both the forward shock wave and the reverse shock wave will be generated in it. Moreover, the forward shock wave and the reverse shock wave do not appear at the same time. The forward shock wave is formed within one second after the supernova explosion, while the reverse shock wave can been seen clearly after about 3 second of explosion.

\section{The variations of $P_s$ and $P_c$ in different SN models}

In the previous section, the density profiles in three supernova models are displayed graphically and we have already known that in the 8.8$M_{\odot}$ model shock wave never exist and in the 18$M_{\odot}$ model only the forward shock wave can be generated, while both the forward and the reverse shock wave are formed in the 10.8$M_{\odot}$ model during the explosion. No matter whether the shock wave exist and no matter how its appearance is, as long as we know the data of density profiles in supernova models, with the theory of neutrino oscillations in matter as formulated in Section 2, we can calculate the survival probability of $\nu_e\rightarrow\nu_e$, as well as the conversion probability of $\nu_x\rightarrow\nu_e$. Given the discrete density values at some points of radius, by substituting the density values into Eq. (4) and through neutrinos' point-by-point evolution, then we can obtain the needed results.

\subsection{The 8.8$M_{\odot}$ model}

In this subsection, based on the theory about the neutrino propagation in matter and applying the density profiles in Section 3, we will show in detail how $P_s$ and $P_c$ change in the 8.8$M_{\odot}$ supernova models, respectively.

In Fig. 4, it is shown how the survival probability of $\nu_e$, $P_s$, changes over time when the neutrino energy takes four different values. Because of the time interval between the density profiles at different time points, $P_s$ changes in a broken-line manner.  As we can see from Fig. 4, for $E$ = 11, 16 and 25 MeV, the value of $P_s$ has a significant reduction from 0.2 s to 0.5 s. After 1 second, $P_s$ is changing all the time, but its changes are very small, and the value of $P_s$ is basically no more than 0.05. While for the line of $E$ = 40 MeV, $P_s$ has small changes before 1 s without dramatic decline, then with the increase of time its value twists and turns up to 0.2 at 6 s.

Fig. 5 details how $P_s$ varies with time and the neutrino energy in a three-dimensional graph. It looks that $P_s$ changes continuously because the curves were processed by "ployfit" under the assumption that there is no sudden change between different time points. This method is also used in other three-dimensional graphs in the following passage. It is found that at the early stage of explosion ($t\leqslant$ 1 s) and when the neutrino energy is relatively small ($E\leqslant$ 10 MeV), the value of $P_s$ is relatively large but declines abruptly in a short time. These obvious changes form some sharp protuberances at the corner of the figure. Then with the increase of time and the neutrino energy, $P_s$ goes into the next stage, in which the variations of $P_s$ are smoother compared with before. Although there are many small fluctuations, the value of $P_s$ always keeps below 0.1 basically in this stage. But as the time is later than 4 s and at high neutrino energy ($E\geqslant$ 40 MeV), some large fluctuations appear, which means $P_s$ is greatly affected by time and the neutrino energy during this period. In parallel with the increase of time and the neutrino energy, $P_s$ is rising with fluctuations in the area of $t\geqslant$ 5 s and $E\geqslant$ 60 MeV, and its value is even close to 0.5 ultimately.

At the same time, we also calculate the conversion probability of $\nu_{x}\rightarrow\nu_{e}$, $P_c$, in the 8.8$M_{\odot}$ model. Fig. 6 shows the time-dependent changes of $P_c$, where the neutrino energy is taken the same values as in Fig. 4. Different from $P_s$, the value of $P_c$ is relatively large. When $t$ = 0.2 s, for $E$ = 11, 25 and 40 MeV, $P_c$ is greater than 0.9; for $E$ = 16 MeV, $P_c$ is greater than 0.75. With time increasing from 0.2 s to 0.5 s, the value of $P_c$ rises up to about 0.97 for all the lines of different neutrino energy. In generally, for $E$ = 11, 16, 25 MeV, $P_c$ always keeps above 0.95 in spite of some small changes after 0.5 s. While for $E$ = 40 MeV, the value of $P_c$ shows a downward trend roughly over the same time period, and decreases to about 0.82 at 6 s.

In Fig. 7, it can be seen how the $P_c$ varies with time and the neutrino energy. At the beginning of explosion ($t\leqslant$ 1 s) and at low neutrino energy ($E\leqslant$ 20 MeV), $P_c$ has abrupt changes so there are some "upside-down peaks" in the corner of this figure. Then the value of $P_c$ goes up and down, and maintains between 0.8 and 1 basically. As time goes on and the neutrino energy gets bigger, in the area of $t\geqslant $ 4 s and $E\geqslant $ 50 MeV, the extent of variations become more and more remarkable.

By comparing Fig. 7 with Fig. 5, it seems that if $P_s$ can be rotated 180 degrees with respect to the XY plane, we would obtain the figure of $P_c$. If so, the sum of $P_s$ and $P_c$ would equal 1, then it means the number of $\nu_ e$ is unchangeable even neutrino oscillations exist in the supernova. If the result of $P_s$ added to $P_c$ is greater than 1, it means the number of $\nu_e$ is more than that when they are emitted. Otherwise, the result of less than 1 implies that the number of $\nu_e$ is less than the original number. In order to find whether the number of $\nu_e$ loses or increases, the results of $P_s$ added to $P_c$ as the function of the neutrino energy are shown in Fig. 8.

Because the result of 0.2 s is quite a bit different from those of other points of time, the results are shown in two figures. In the left figure, it is found that when $t$ = 0.2 s, the number of $\nu_e$ is severely changeable with the neutrino energy, which means at the beginning of explosion, the conversion between $\nu_e$ and $\nu_x$ is very drastic. In the right figure, we can see the results are always changing but trend toward stability compared with the case of 0.2 s and keep within the range of 0.93 - 1.07. That is, the losing and increasing ratio of $\nu_e$ are both no more than 7$\%$ during 0.5 s - 6 s of explosion.

\subsection{The 10.8$M_{\odot}$ model}

In Fig. 9, there are four typical examples of $P_s$ at different neutrino energies changing with time in the 10.8$M_{\odot}$ model. It is found that the value of $P_s$ is less than 0.2 at 0.5 s for $E$ = 11, 16, 25, 40 MeV. When $t$ = 1 s, $P_s$ of four different neutrino energies all drop to about 0.02 and keep the value till $8s$. As for $E$ =11, 16, 25 MeV, $P_s$ maintains its value until 10 s; but for $E$ = 40 MeV, at 9 s $P_s$ increases to about 0.08 and then falls back to approach the previous value before 8 s. Compared with Fig. 4, it seems that the time-varying fluctuations of $P_s$ in the 10.8$M_{\odot}$ model are much less than those in the 8.8$M_{\odot}$ model.

The variations of $P_s$ with time and the neutrino energy in the 10.8$M_{\odot}$ model is shown by a three-dimensional graph in Fig. 10. At 0.5 s, the value of $P_s$ changes significantly with the increase of the neutrino energy, so there are many fluctuations along the axis of the neutrino energy. From $t$ = 0.5 s to $t$ = 1 s, $P_s$ has an abrupt decrease, which is in accord with the case shown in Fig. 9. With time going on, despite the neutrino energy is different, $P_s$ goes into a relatively stable state and its value is always around zero. But in the region of $t\geqslant$ 8 s and $E\geqslant$ 40 MeV, some obvious fluctuations of $P_s$ revive again, however, its value never exceeds 0.2.

Overall, there is an obvious difference between the 10.8$M_{\odot}$ model and the 8.8$M_{\odot}$ model about $P_s$. That is, $P_s$ is more stable in the 10.8$M_{\odot}$ model than in the 8.8$M_{\odot}$ model. At the same time, two models also have some similarities: at the beginning of explosion ($t\leqslant$ 1 s), $P_s$ has abrupt changes in a short time and fluctuates significantly with the neutrino energy; in the area of large neutrino energy and during the later time of explosion, fluctuations of $P_s$ become obvious and frequent.

In Fig. 11, it can be seen how $P_c$ changes with time when the neutrino energy takes four different values in the 10.8$M_{\odot}$ model. Obviously, for all the lines the value of $P_c$ increases from more than 0.82 of 0.5 s to about 0.98 of 1 s. Afterwards, for $E$ = 11, 16 MeV, $P_c$ keeps invariable basically from 2 s to 9 s and decreases slightly at 10 s. On the lines of $E$ = 25, 40 MeV, there are relatively small drops at 3 s and relatively large drops at 9 s respectively, yet without significant changes at other points of time.

The variations of $P_c$ with time and the neutrino energy in the 10.8$M_{\odot}$ model is shown in Fig. 12. Similar to $P_s$ in Fig. 10, at 0.5 s, $P_c$ has many fluctuations along the axis of the neutrino energy and a sudden change from 0.5 s to 1 s. When $t\geqslant$ 8 s and $E\geqslant$ 40 MeV, the fluctuations of $P_c$ exist again, and its value varies between 0.8 and 1 approximately. While in other areas, $P_c$ maintains the basic stability. Generally, the figure of $P_c$ in the 10.8$M_{\odot}$ model is  smoother by comparison with $P_c$ of the 8.8$M_{\odot}$ model shown in Fig. 7 .

Likewise, we also plot the results of $P_s$ added to $P_c$ to find how the number of $\nu_e$ changes in the 10.8$M_{\odot}$ model. Same as Fig. 8, there are two figures in Fig. 13. In the left figure, it can be seen that at 0.5 s, the results hover drastically around 1 with the increase of the neutrino energy, which means at this point of time, the conversion of $\nu_e$ and $\nu_x$ is very dramatic and greatly affected by the neutrino energy. By contrast, the results of other points of time are relatively stable. The right figure indicate that from 1 s to 10 s of the explosion, the sum of $P_s$ and $P_c$ fluctuates in the range of 0.96 and 1.01. At 1 s, the results exceed 1 occasionally, while from 2 s to 10 s, the results are almost under the lever of 1, which illustrates that the number of $\nu_e$ is more likely to lose during this period. The losing rate and the increasing rate of $\nu_e$ is less than 4$\%$ and 1$\%$, respectively.

\subsection{The 18$M_{\odot}$ model}

In Fig. 14, it is illustrated how $P_s$ changes as the function of time at four different neutrino energies in the 18$M_{\odot}$ model. In this figure, we can see that at 1 s, for the four neutrino energies, $P_s$ drops below 0.05 from its original value, and then it remains so until the end of time.

In Fig. 15, it is found that $P_s$ has many prominent fluctuations at 0.5 s with the change of the neutrino energy. After 1 s, no change can be observed so the value of $P_s$ almost becomes a plane in the three-dimensional graph. This phenomenon corresponds to the straight lines from 1 s to 10 s in Fig. 13. The value of $P_s$ is very close to 0, meaning that $\nu_e$ almost completely convert into $\nu_x$ during this period.

In Fig. 16, for the four lines of different neutrino energies, the value of $P_c$ in the 18$M_{\odot}$ model has an obvious rise at 1 s and then basically remains constant although the time is changing. The only noticeable one is a slight decline at 1 s for $E$ = 11 and 16 MeV.

In Fig. 17, it can be seen that $P_c$ changes with the neutrino energy very significantly at 0.5 s. Just like Fig. 7 and Fig. 12, there are many fluctuations on the edge of the figure. From 0.5 s to 1 s, the value of $P_c$ goes up obviously, which can be seen more clearly in Fig. 16. After the marked change, $P_c$ is nearly invariable and remains broadly flat over the rest of the region consisting of time and the neutrino energy. And the value of $P_c$ always near 1 suggests that almost all $\nu_x$ convert into $\nu_e$ after 1 second of explosion.

In the left picture of Fig. 18, the line of 0.5 s varies with the neutrino energy remarkably. But on other lines, it is difficult to observe significant changes. In the right figure, it is shown that the variations of results at low energies ($E\leqslant$ 30 MeV) are a little remarkable than those at high energies during the time of 1 s - 10 s. But the change interval is always between 0.975 and 1.01. In other words, the permissible losing rate of $\nu_e$ is no more than 2.5$\%$ and the increasing rate is no more that 1$\%$ during this time.

\section{Summery and Discussions}

By comparing the variations of $P_s$ in three models, it is observed that at the start of explosion ($t<$ 1 s), $P_s$ changes abruptly in a short time and changes significantly with the neutrino energy. Later on, it remains relatively stable. But with the increase of time and the neutrino energy, fluctuations of $P_s$ start to get more and more prominent in the 8.8$M_{\odot}$ model and the 10.8$M_{\odot}$ model. Meanwhile, the intensity and the scale of fluctuations in the 10.8$M_{\odot}$ model are lower and smaller than those in the 8.8$M_{\odot}$ model. Nevertheless, this "restart-fluctuating" phenomenon does not appear in the 18$M_{\odot}$ model, in consequence the previous stable state is maintained until the end. In brief, $P_s$ is getting more stable with the increase of supernova masses in the models.

Similar to the variations of $P_s$, $P_c$ in three models also have an abrupt change before 1 s and can be affected obviously by the neutrino energy at the initial stage of explosion. Followed by a relatively stable phase, $P_c$ restarts to fluctuate over time and with the increase of the neutrino energy. Same as the case of $P_s$, with the increase of supernova mass in the model, the relative stable area of $P_c$ is getting bigger and bigger.

By plotting the results of $P_s$ added to $P_c$ in three models, we can analyze how the number of $\nu_e$ changes with the neutrino energy at different times. It turns out that the number of $\nu_e$ is extremely unstable at the beginning of explosion ($t<$ 1 s), reflecting the dramatic conversions between $\nu_e$ and $\nu_x$. During the next period of explosion (1 s - 6 s for the 8.8$M_{\odot}$ model; 1 s - 10 s for the 10.8$M_{\odot}$ and the 18$M_{\odot}$ model), the number of $\nu_e$ has some small changes with time and the neutrino energy. Generally, if we could detect $\nu_e$ after one second of explosion outside the supernova, compared with the moment when $\nu_e$ is just emitted, the number of $\nu_e$ might be increasing or reduced. Its change rate is no more than 7$\%$, 4$\%$ and 2.5$\%$ in the 8.8$M_{\odot}$, 10.8$M_{\odot}$ and 18$M_{\odot}$ models, respectively. It seems that the number of $\nu_e$ is more stable in the more massive supernova, although the conversions between $\nu_e$ and $\nu_x$ are very drastic.

In summary, the supernova mass in the model determines the density profiles in the supernova. Consequently, it can affect the interactions between neutrinos and matter, and then make changes to $P_s$, $P_c$ and the number of neutrinos that can be detected. So it needs to be stressed that the supernova mass is an important factor when properties of supernova neutrinos are studied.

There is no doubt that if we want to obtain the number of neutrinos that can be detected on the earth, discussing the shock effects is merely not enough. In Refs. \cite{Guo}\cite{Huang}, we have already studied on the earth effects comprehensively and the relevant researches on the collective effects are ongoing currently \cite{Dasgupta}-\cite{Hu}. So far, we have already known that the collective effects are very important on the conversions between different neutrinos and affect other aspects about neutrinos, such as the mass hierarchy and luminosity \cite{Chakraborty1}\cite{Fogli02}. With the progress of study on the impacts of shock wave effects and collective effects, it is possible to obtain the results in good agreement with experiments.

\section{Acknowledgments}

We are very grateful to Tobias Fisher for providing the simulation data about the density profiles of supernovae. This work was supported in part by National Science Foundation of China (Project Numbers 11175020, 11275025).


\newpage

\begin{figure}
\includegraphics[width=0.67\textwidth]{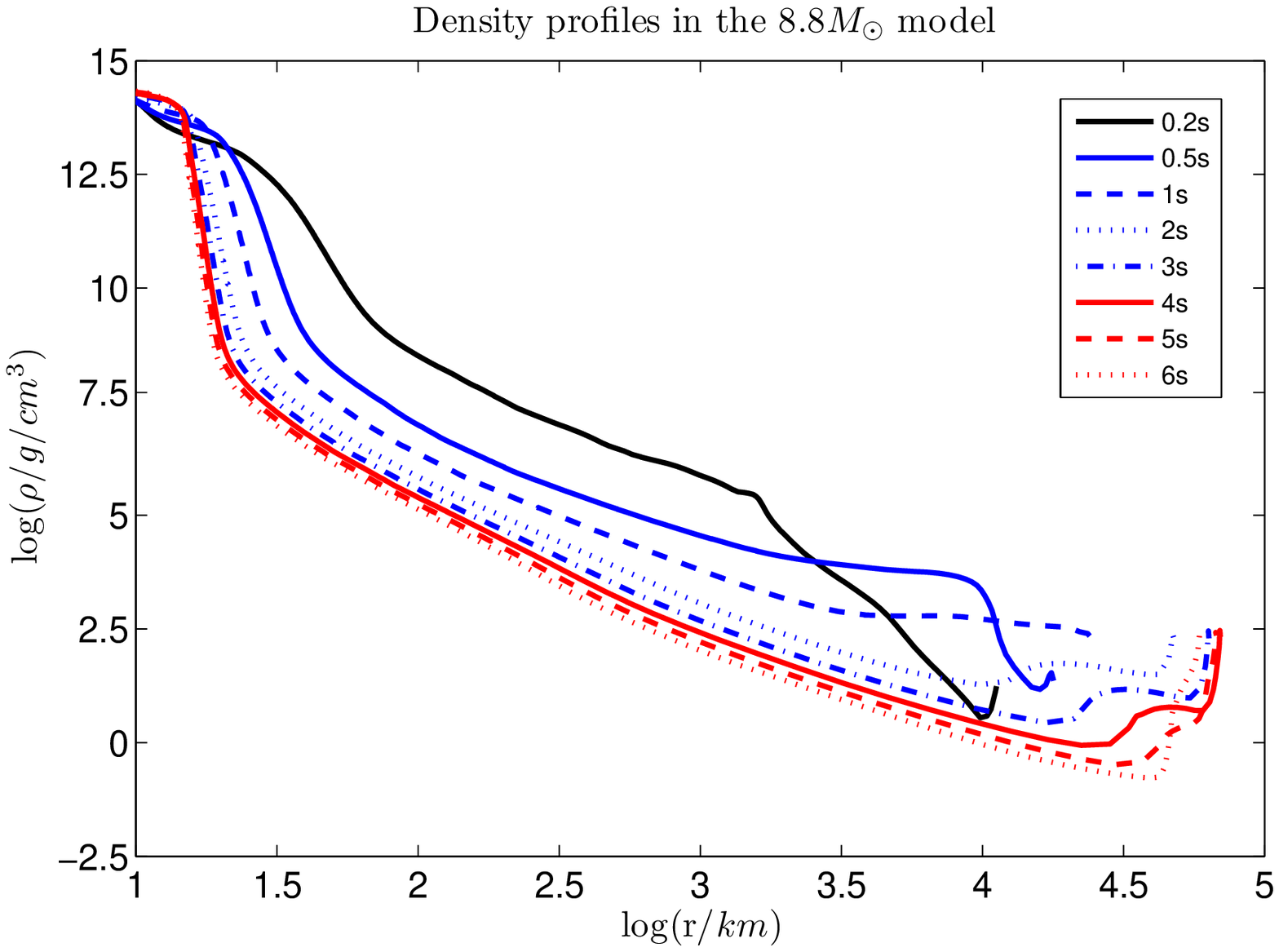}\\
\centerline{Fig. 1}
\caption{Density profiles of different times in the 8.8$M_{\odot}$ model. The dark solid curve corresponds to 0.2 s. The blue solid, dashed, dotted and dot-dashed curves correspond to 0.5 s, 1 s, 2 s and 3 s, respectively. The red solid, dashed and dotted curves correspond to 4 s, 5 s and 6 s, respectively.}
\label{fig:1}  
\end{figure}

\begin{figure}
\includegraphics[width=0.6\textwidth]{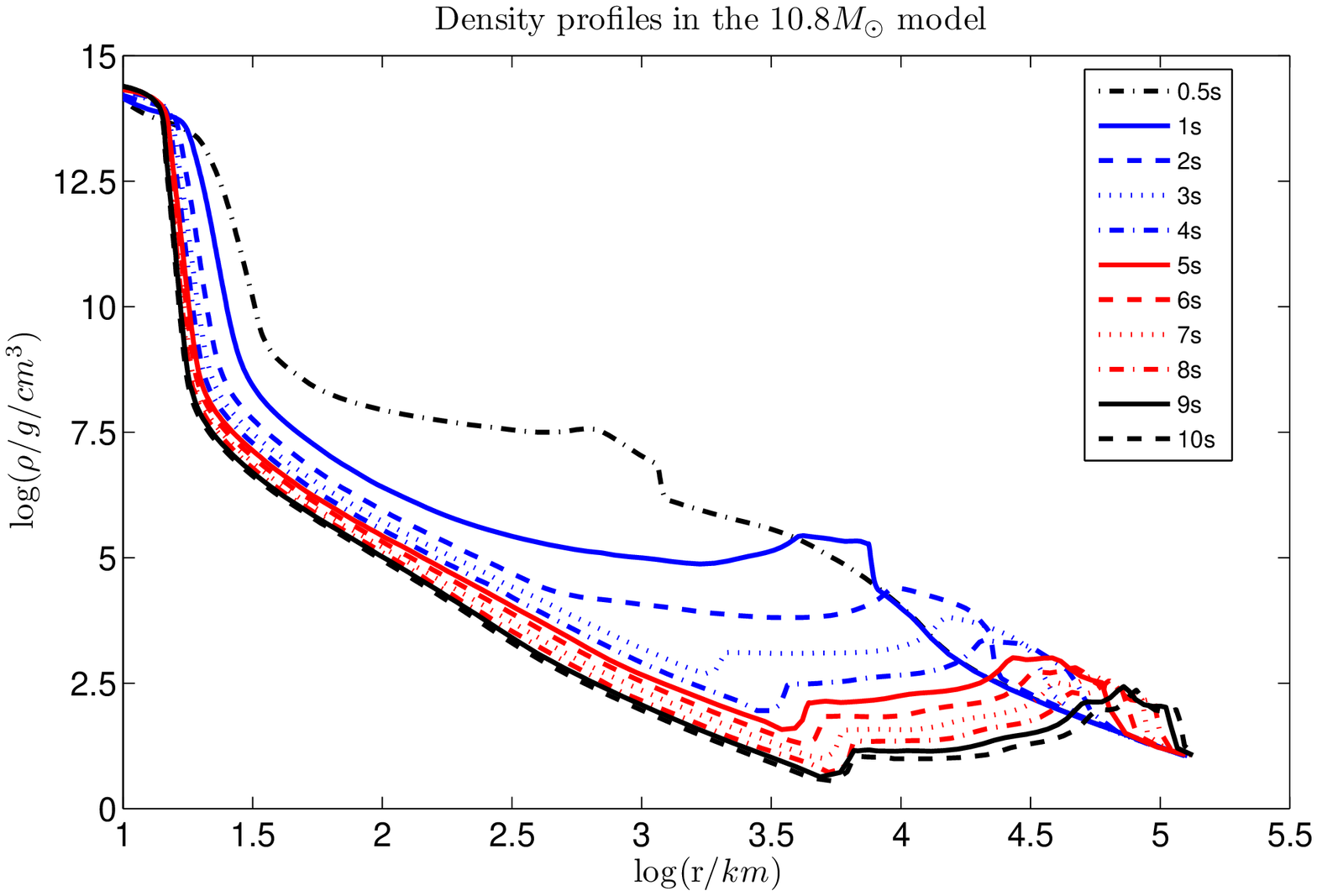}\\
\centerline{Fig. 2}
\caption{Density profiles of different times in the 10.8$M_{\odot}$ model. The dark dot-dashed curve corresponds to 0.5 s. The blue solid, dashed, dotted and dot-dashed curves correspond to 1 s, 2 s, 3 s and 4 s, respectively. The red solid, dashed, dotted and dot-dashed curves correspond to 5 s, 6 s, 7 s and 8 s, respectively. The dark solid and dashed curves correspond to 9 s and 10 s, respectively.}
\end{figure}

\begin{figure}
\includegraphics[width=0.63\textwidth]{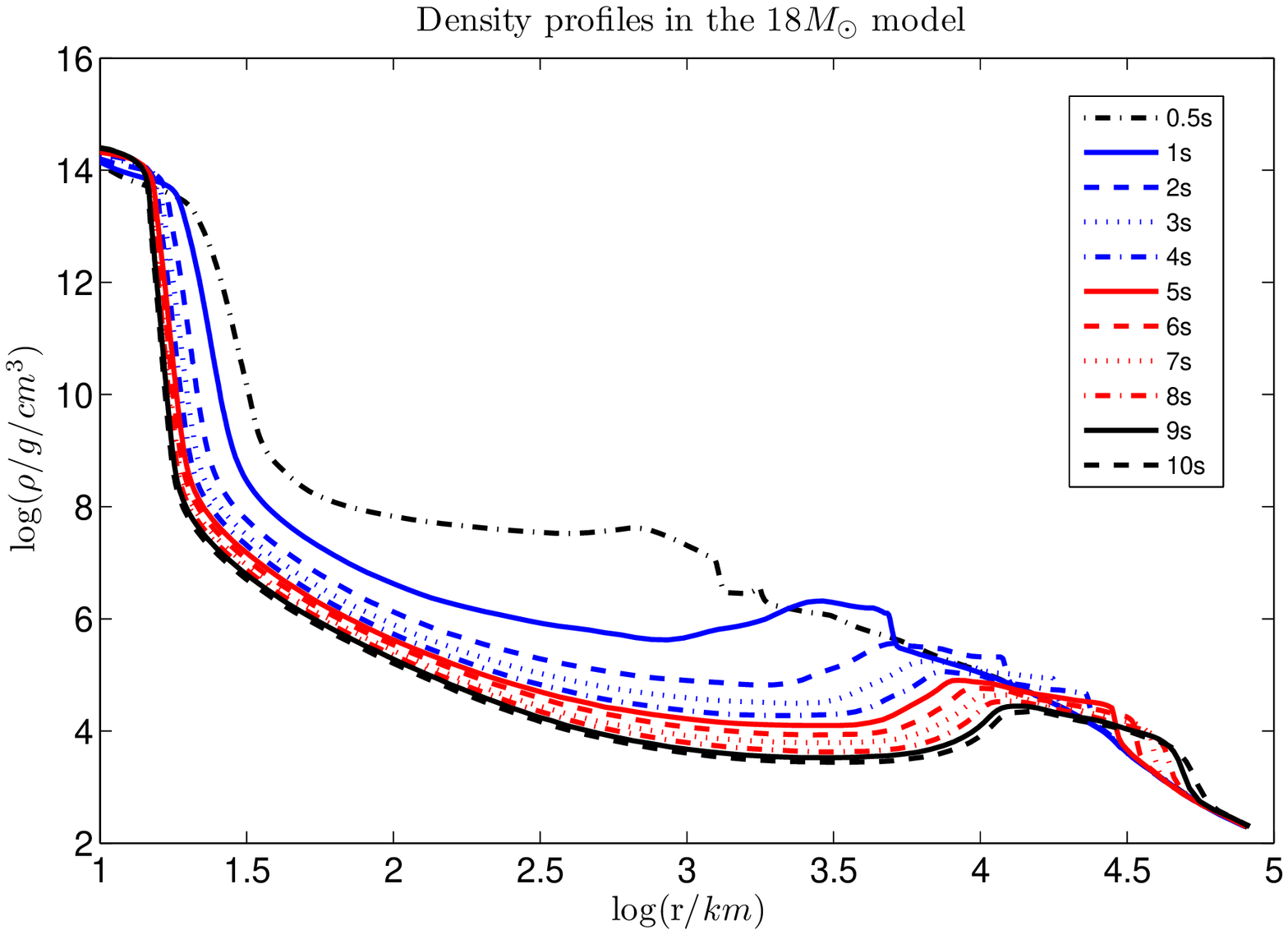}\\
\centerline{Fig. 3}
\caption{Density profiles of different times in the 18$M_{\odot}$ model. The dark dot-dashed curve corresponds to 0.5 s. The blue solid, dashed, dotted and dot-dashed curves correspond to 1 s, 2 s, 3 s and 4 s, respectively. The red solid, dashed, dotted and dot-dashed curves correspond to 5 s, 6 s, 7 s and 8 s, respectively. The dark solid and dashed curves correspond to 9 s and 10 s, respectively.}
\end{figure}

\begin{figure}
\includegraphics[width=0.6\textwidth]{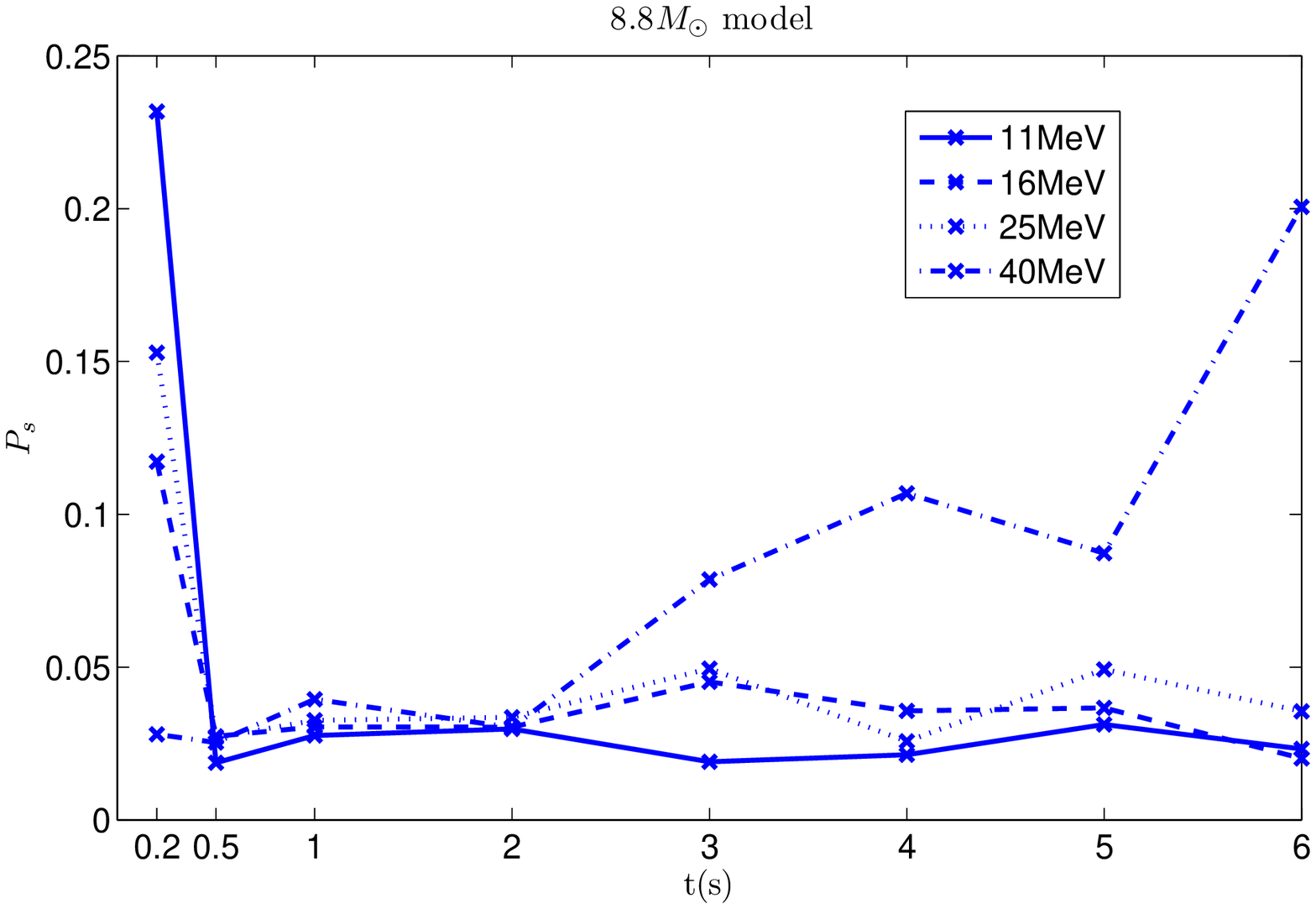}\\
\centerline{Fig. 4}
\caption{$P_s$ as a function of the time $t$ for four different neutrino energies in the 8.8$M_{\odot}$ model. The solid, dashed, dotted and dot-dashed curves correspond to $E$ = 11, 16, 25 and 40 MeV, respectively. The cross points are obtained from our calculations.}
\end{figure}

\begin{figure}
\includegraphics[width=0.6\textwidth]{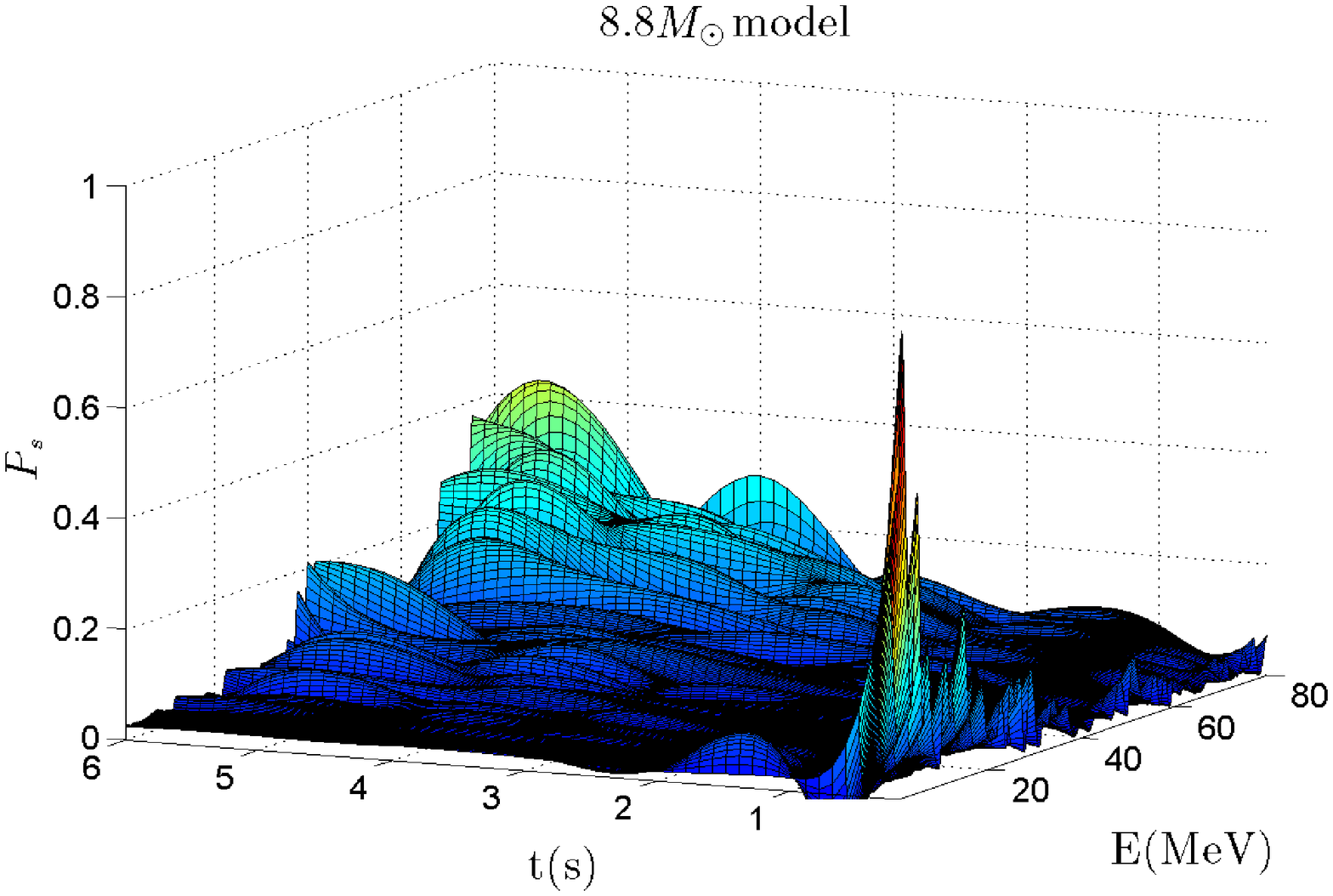}\\
\centerline{Fig. 5}
\caption{The variations of $P_s$ with time and the neutrino energy in the 8.8$M_{\odot}$ model. The scopes of independent variables are: 0.2 s $\leq t\leq $ 6 s, 1 MeV $\leq  E\leq $ 80 MeV, respectively.}
\end{figure}

\begin{figure}
\includegraphics[width=0.6\textwidth]{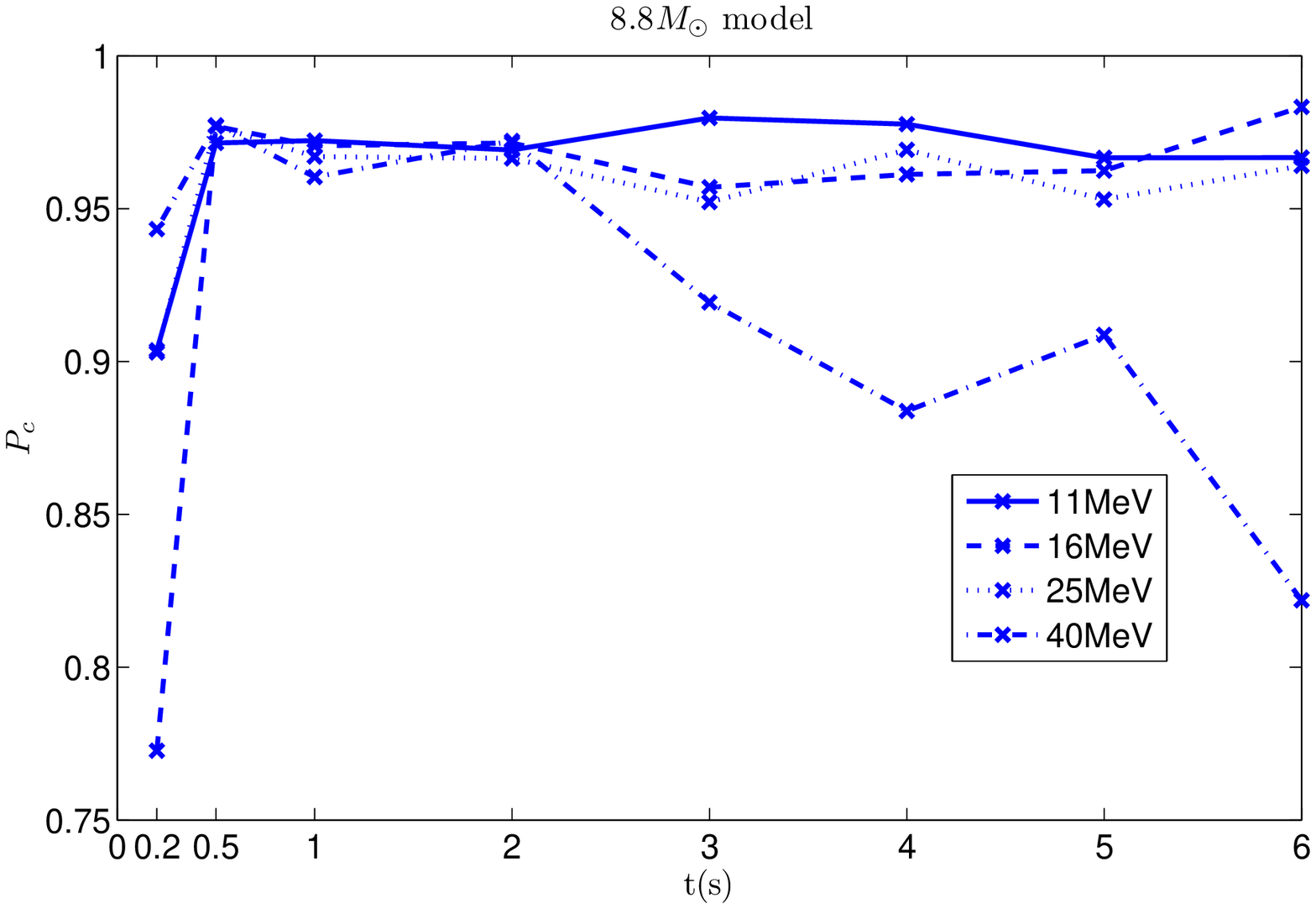}\\
\centerline{Fig. 6}
\caption{$P_c$ as a function of the time $t$ for four different neutrino energies in the 8.8$M_{\odot}$ model. The solid, dashed, dotted and dot-dashed curves correspond to $E$ = 11, 16, 25 and 40 MeV, respectively. The cross points are obtained from our calculations.}
\end{figure}

\begin{figure}
\includegraphics[width=0.6\textwidth]{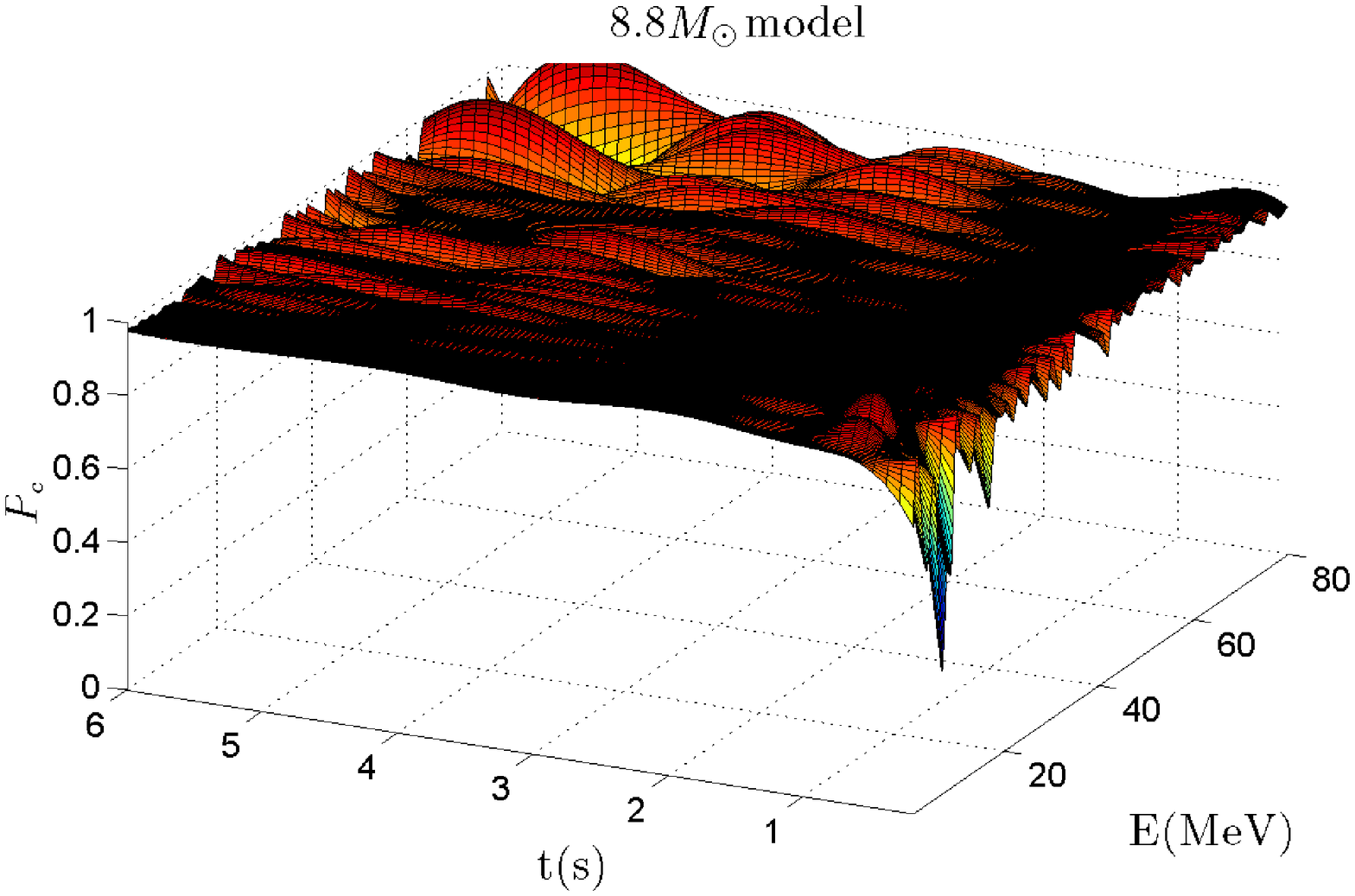}\\
\centerline{Fig. 7}
\caption{The variations of $P_c$ with time and the neutrino energy in the 8.8$M_{\odot}$ model. The scopes of independent variables are: 0.2 s $\leq t\leq$ 6 s, 1 MeV $\leq E\leq$ 80 MeV, respectively.}
\end{figure}

\begin{figure}
\includegraphics[width=0.6\textwidth]{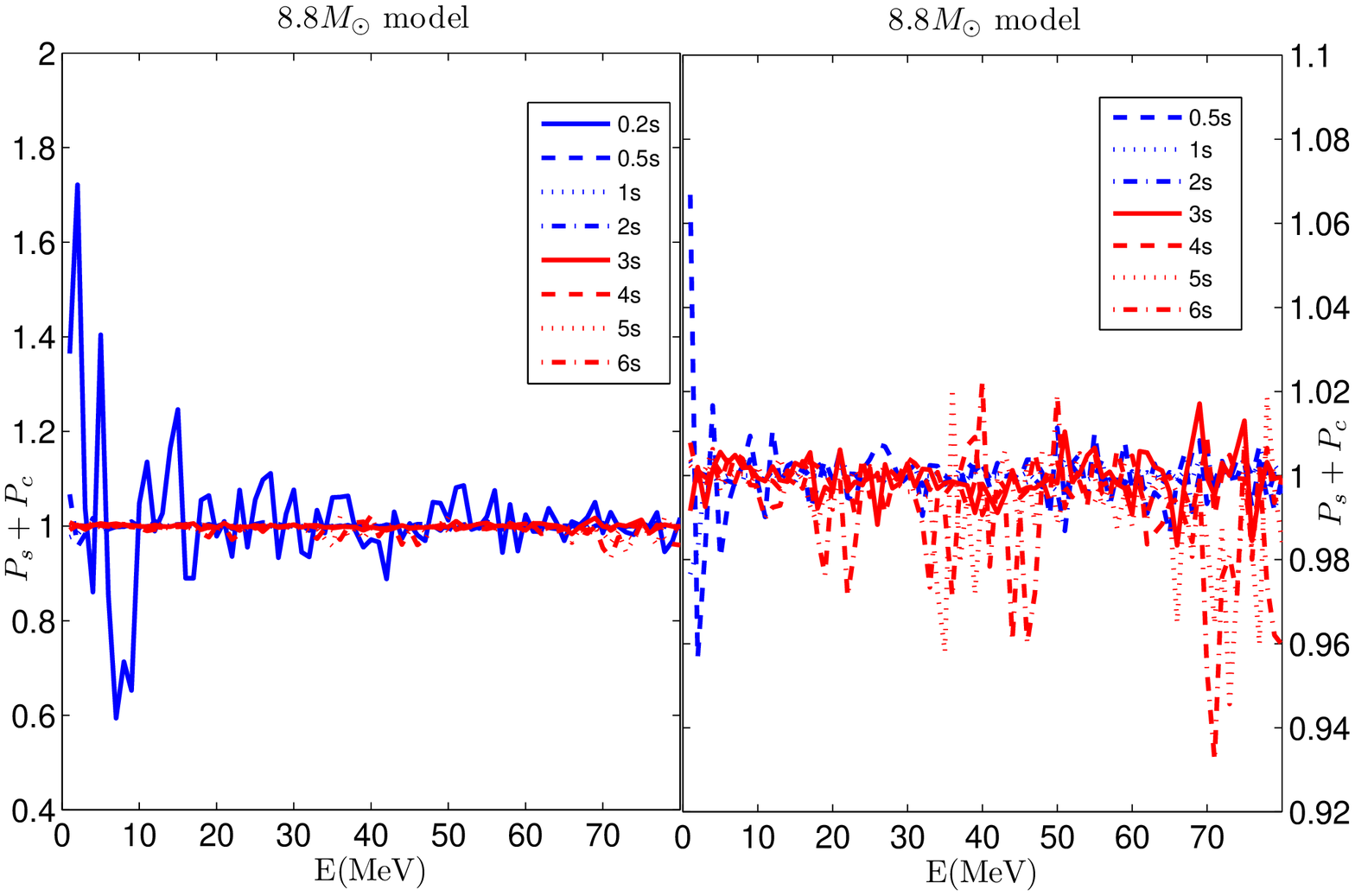}\\
\centerline{Fig. 8}
\caption{The results of $P_s$ added to $P_c$ as a function of the neutrino energy at different times in the 8.8$M_{\odot}$ model. The blue solid, dashed, dotted and dot-dashed curves correspond to 0.2 s, 0.5 s, 1 s and 2 s, respectively. The red solid, dashed, dotted and dot-dashed curves correspond to 3 s, 4 s, 5 s and 6 s, respectively. In the left figure, there are 8 curves of all different times. In the right figure, there are 7 curves except the curve of 0.2 s.}
\end{figure}

\begin{figure}
\includegraphics[width=0.6\textwidth]{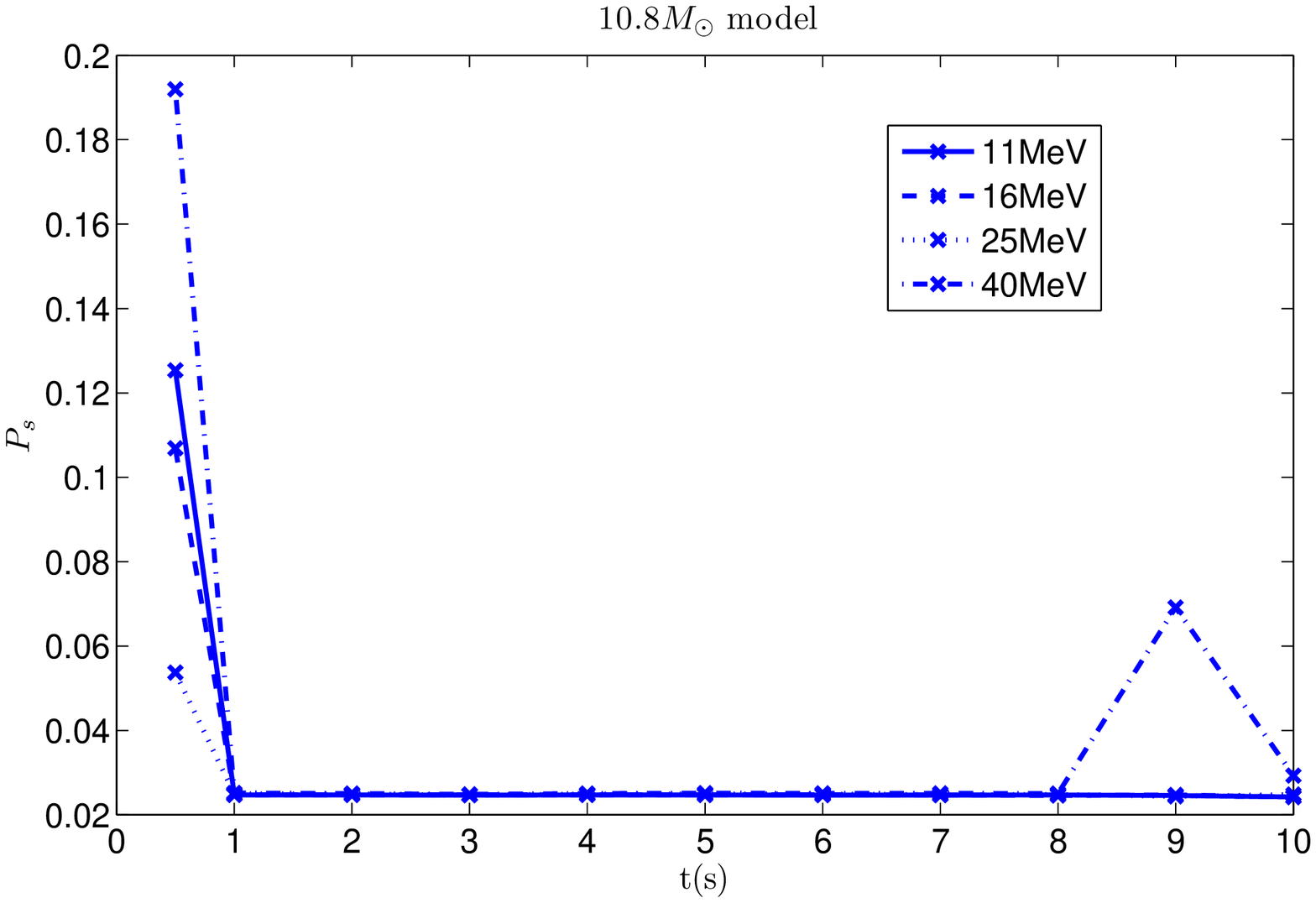}\\
\centerline{Fig. 9}
\caption{$P_s$ as a function of the time $t$ for four different neutrino energies in the 10.8$M_{\odot}$ model. The solid, dashed, dotted and dot-dashed curves correspond to $E$ = 11, 16, 25 and 40 MeV, respectively. The cross points are obtained from our calculations.}
\end{figure}

\begin{figure}
\includegraphics[width=0.6\textwidth]{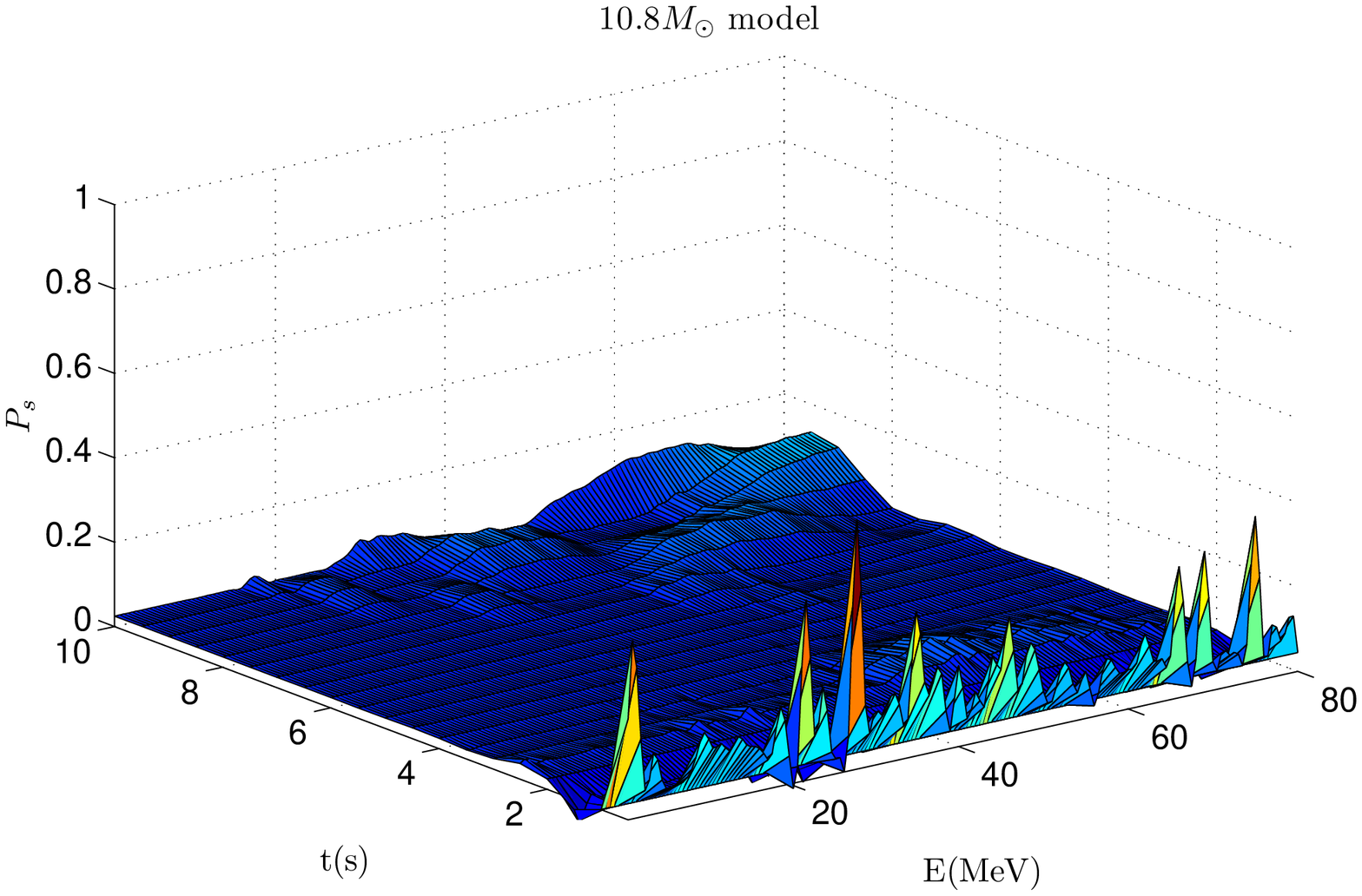}\\
\centerline{Fig. 10}
\caption{The variations of $P_s$ with time and the neutrino energy in the 10.8$M_{\odot}$ model. The scopes of independent variables are: 0.5 s $\leq  t\leq $ 10 s, 1 MeV $\leq E\leq$ 80 MeV, respectively.}
\end{figure}

\begin{figure}
\includegraphics[width=0.6\textwidth]{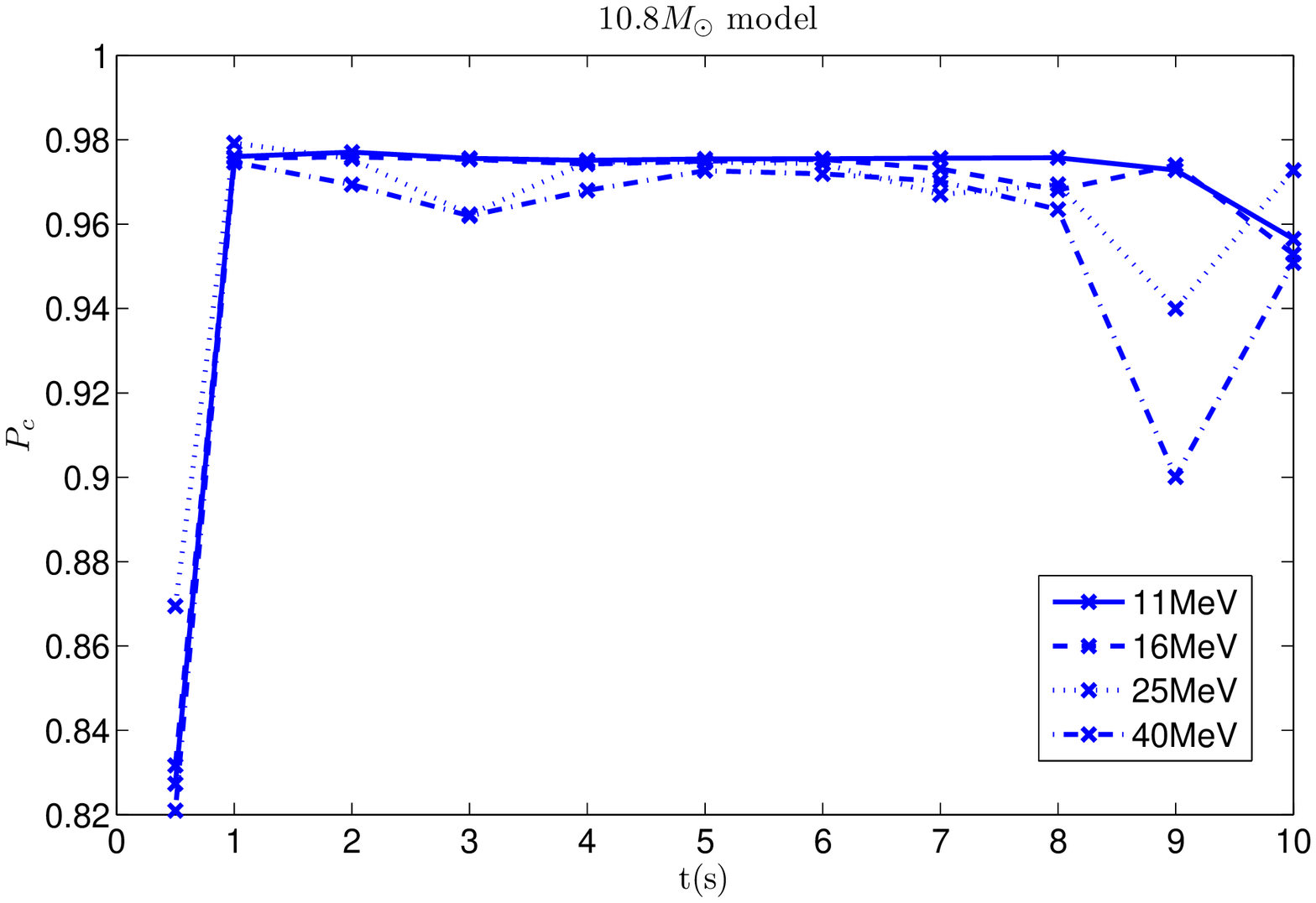}\\
\centerline{Fig. 11}
\caption{$P_c$ as a function of the time $t$ for four different neutrino energies in the 10.8$M_{\odot}$ model. The solid, dashed, dotted and dot-dashed curves correspond to $E$ = 11, 16, 25 and 40 MeV, respectively. The cross points are obtained from our calculations.}
\end{figure}

\begin{figure}
\includegraphics[width=0.6\textwidth]{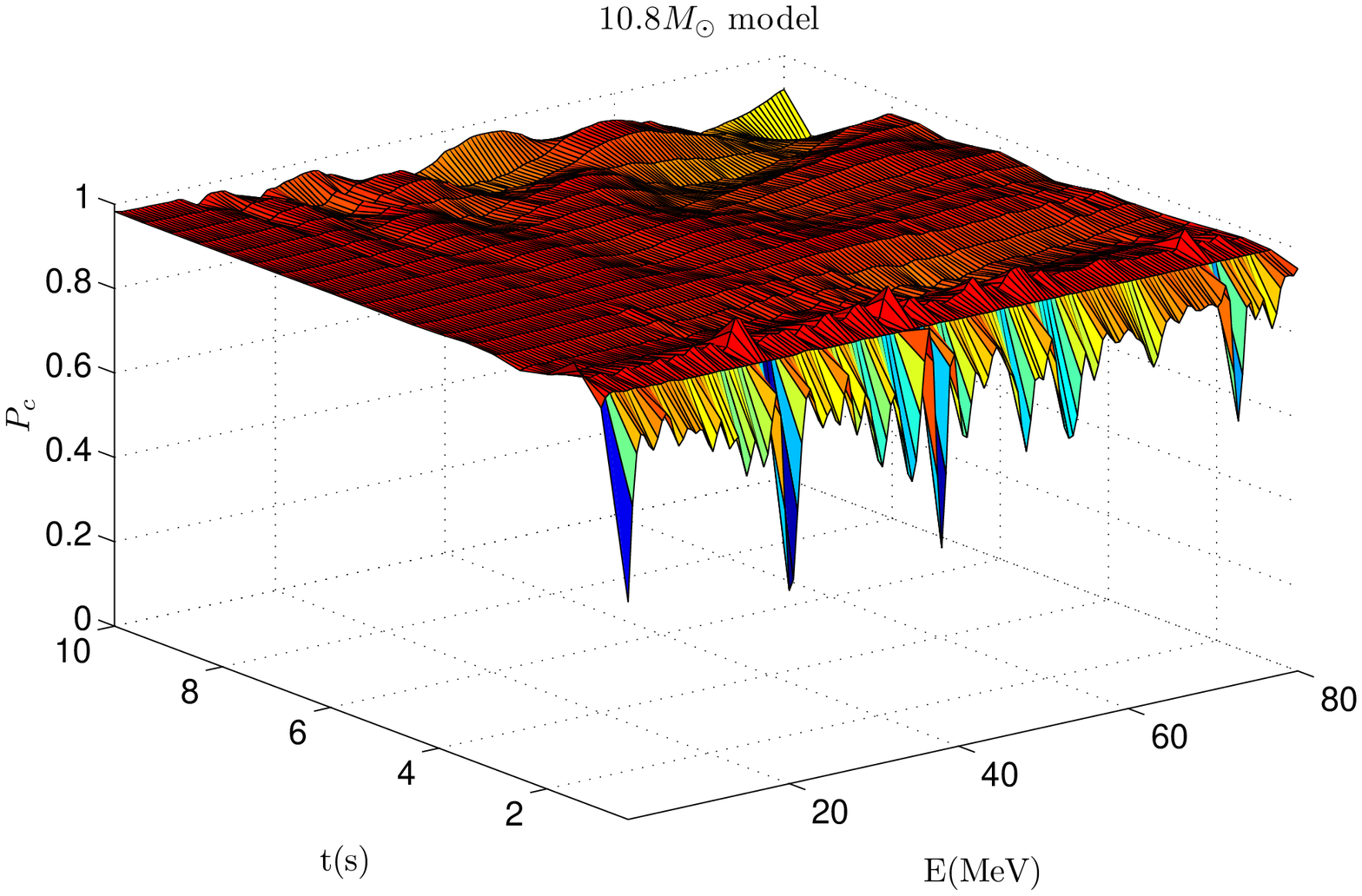}\\
\centerline{Fig. 12}
\caption{The variations of $P_c$ with time and the neutrino energy in the 10.8$M_{\odot}$ model. The scopes of independent variables are: 0.5 s $\leq  t\leq $ 10 s, 1 MeV $\leq  E\leq $ 80 MeV, respectively.}
\end{figure}

\begin{figure}
\includegraphics[width=0.6\textwidth]{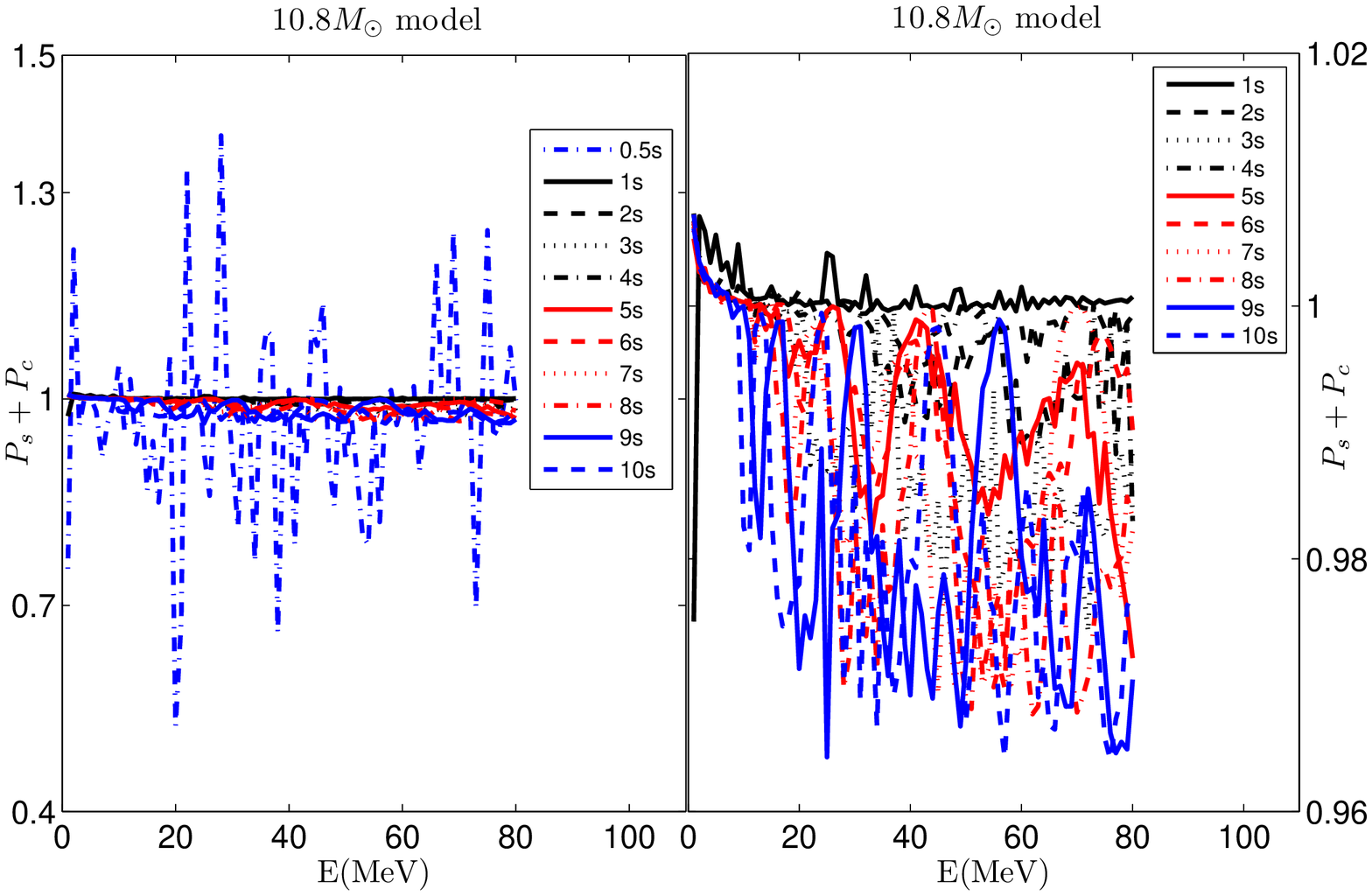}\\
\centerline{Fig. 13}
\caption{The results of $P_s$ added to $P_c$ as a function of the neutrino energy at different times in the 10.8$M_{\odot}$ model. The blue dot-dashed curve correspond to 0.5 s. The dark solid, dashed, dotted and dot-dashed curves correspond to 1 s, 2 s, 3 s and 4 s, respectively. The red solid, dashed, dotted and dot-dashed curves correspond to 5 s, 6 s, 7 s and 8 s, respectively. The blue solid and dashed curves correspond to 9 s and 10 s, respectively. In the left figure, there are 11 curves of all different times. In the right figure, there are 10 curves except the curve of 0.5 s.}
\end{figure}

\begin{figure}
\includegraphics[width=0.6\textwidth]{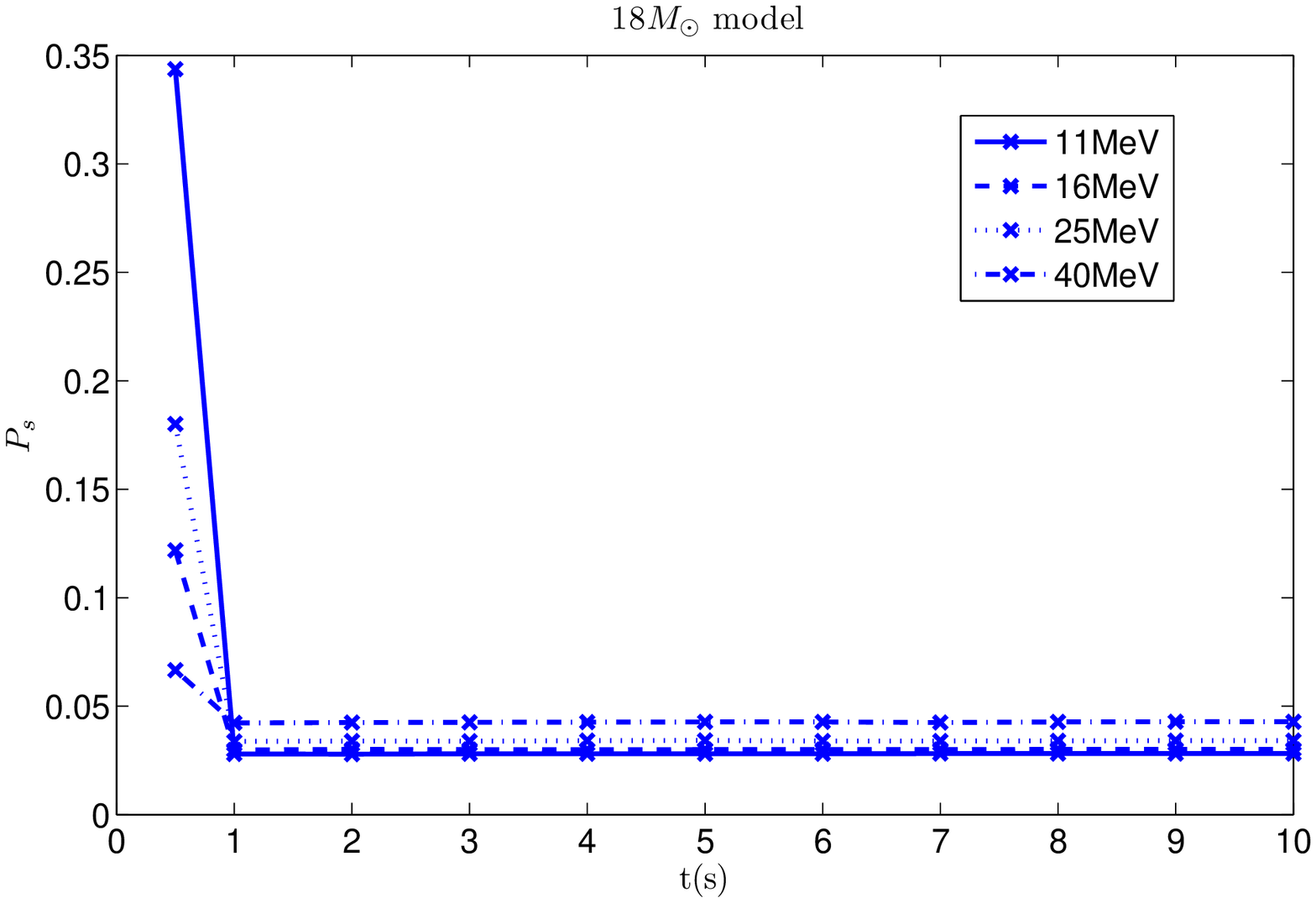}\\
\centerline{Fig. 14}
\caption{$P_s$ as a function of the time $t$ for four different neutrino energies in the 18$M_{\odot}$ model. The solid, dashed, dotted and dot-dashed curves correspond to $E$ = 11, 16, 25 and 40 MeV, respectively. The cross points are obtained from our calculations.}
\end{figure}

\begin{figure}
\includegraphics[width=0.6\textwidth]{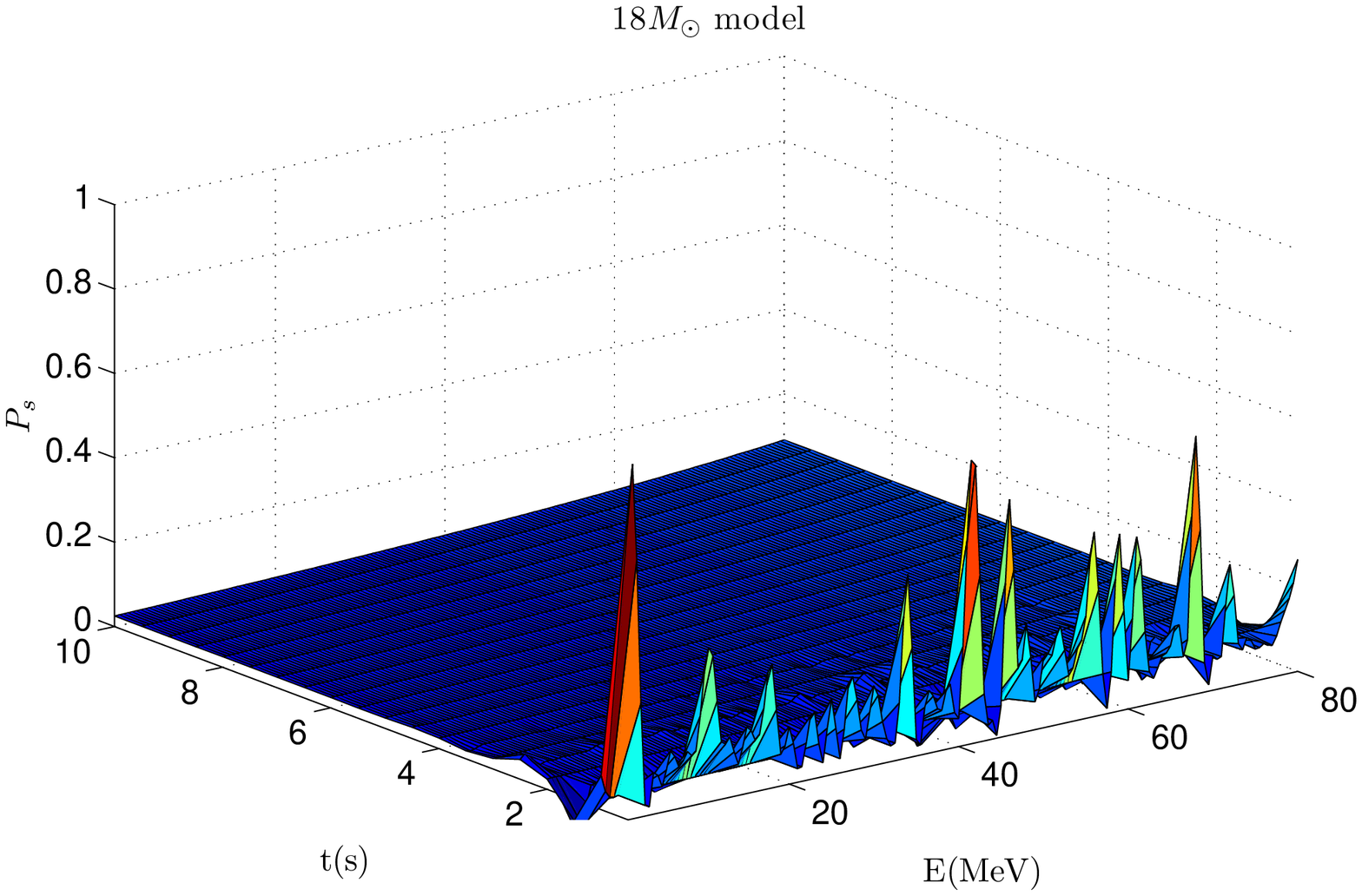}\\
\centerline{Fig. 15}
\caption{The variations of $P_s$ with time and the neutrino energy in the 18$M_{\odot}$ model. The scopes of independent variables are: 0.5 s$\leq  t\leq$ 10 s, 1 MeV $\leq  E\leq$ 80 MeV, respectively.}
\end{figure}

\begin{figure}
\includegraphics[width=0.6\textwidth]{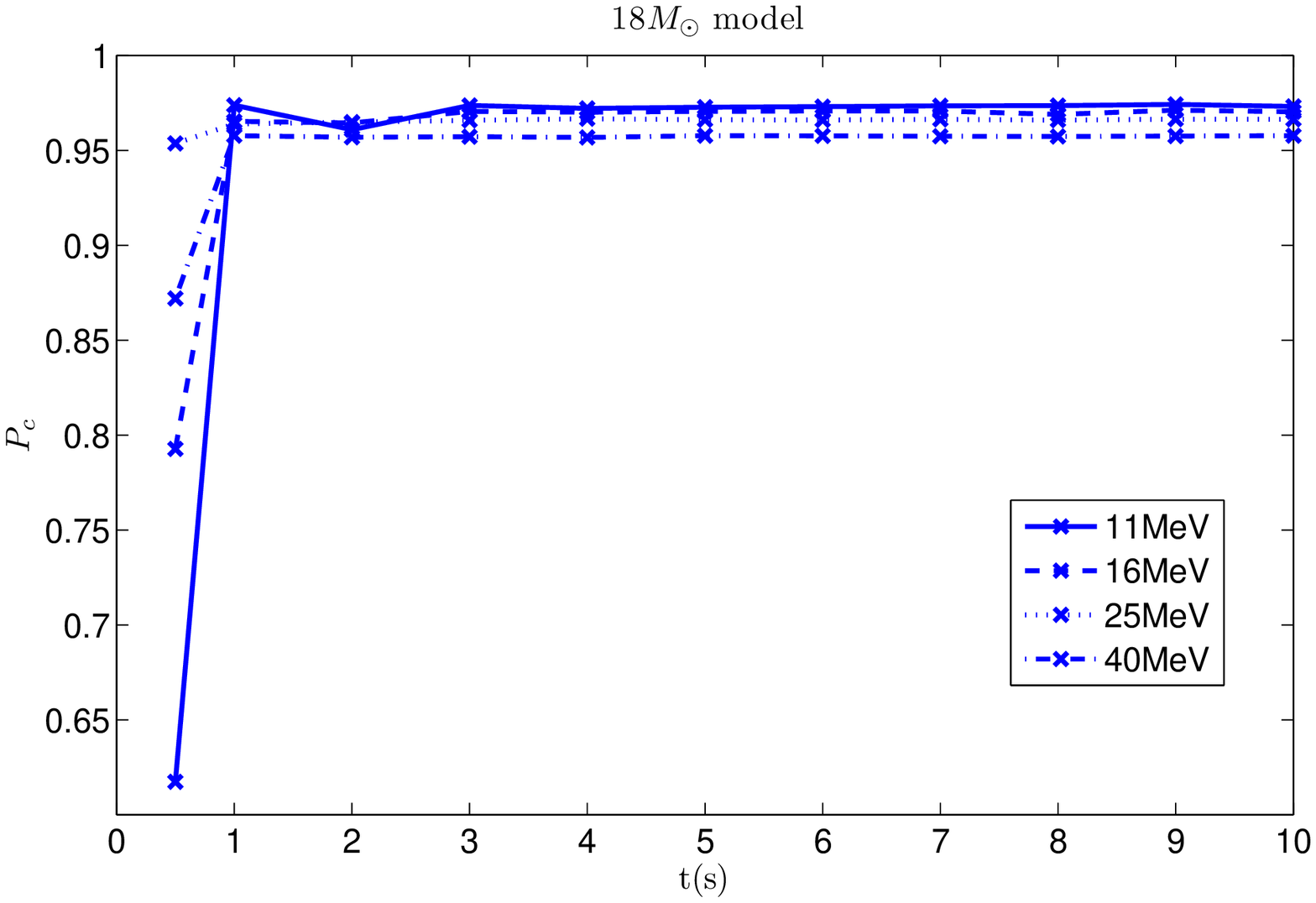}\\
\centerline{Fig. 16}
\caption{$P_c$ as a function of the time $t$ for four different neutrino energies in the 18$M_{\odot}$ model. The solid, dashed, dotted and dot-dashed curves correspond to $E$ = 11, 16, 25 and 40 MeV, respectively. The cross points are obtained from our calculations.}
\end{figure}

\begin{figure}
\includegraphics[width=0.6\textwidth]{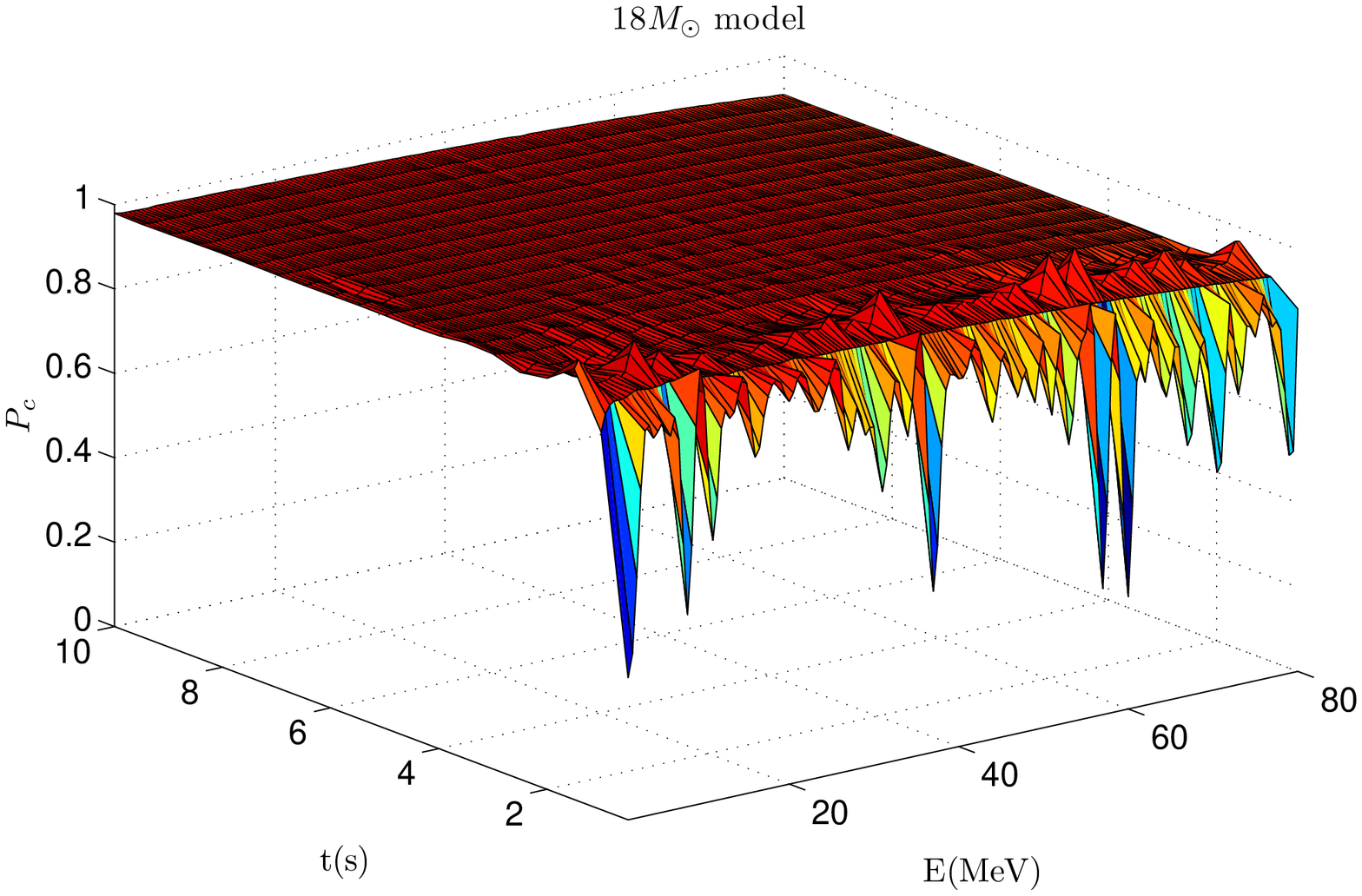}\\
\centerline{Fig. 17}
\caption{The variations of $P_c$ with time and the neutrino energy in the 18$M_{\odot}$ model. The scopes of independent variables are: 0.5 s $\leq  t\leq$ 10 s, 1 MeV $\leq  E\leq$ 80 MeV, respectively.}
\end{figure}

\begin{figure}
\includegraphics[width=0.6\textwidth]{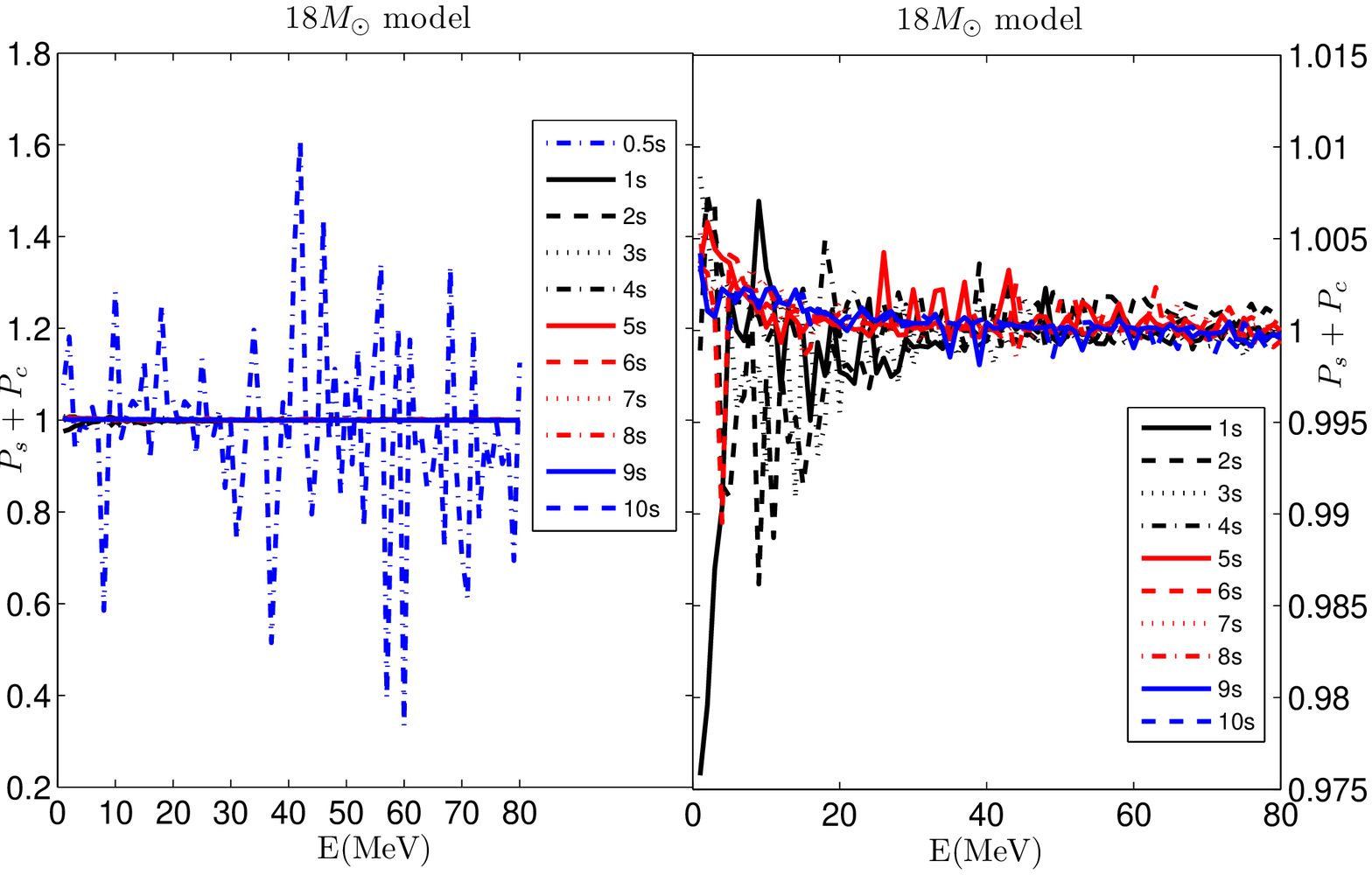}\\
\centerline{Fig. 18}
\caption{The results of $P_s$ added to $P_c$ as a function of the neutrino energy at different times in the 18$M_{\odot}$ model. The blue dot-dashed curve correspond to 0.5 s. The dark solid, dashed, dotted and dot-dashed curves correspond to 1 s, 2 s, 3 s and 4 s, respectively. The red solid, dashed, dotted and dot-dashed curves correspond to 5 s, 6 s, 7 s and 8 s, respectively. The blue solid and dashed curves correspond to 9 s and 10 s, respectively. In the left figure, there are 11 curves of all different times. In the right figure, there are 10 curves except the curve of 0.5 s.}
\end{figure}

\cleardoublepage


\end{spacing}
\end{document}